\documentclass[conference]{IEEEtran}
\usepackage{graphicx}
\usepackage[belowskip=-15pt,aboveskip=0pt]{caption}
\usepackage{amsmath}
\usepackage{multicol}
\usepackage{color}
\usepackage{amssymb}
\usepackage{subfig}
\usepackage{cite}

\begin{document}
\title{Nobody But You: Sensor Selection for Voltage Regulation in Smart Grid}
\author{\IEEEauthorblockN{Rukun Mao and Husheng Li}
\IEEEauthorblockA{Department of Electrical Engineering and Computer Science\\
The University of Tennessee at Knoxville\\
Email: \{rmao,husheng\}@eecs.utk.edu}} 
\maketitle
\begin{abstract}
	The increasing availability of distributed energy resources (DERs) and sensors in smart grid, as well as overlaying communication network, provides
	substantial potential benefits for improving the power system's reliability.
	In this paper, the problem of sensor selection is studied for the MAC layer design of wireless sensor networks for regulating the voltages in smart grid.
	The framework of hybrid dynamical system is proposed, using Kalman filter for voltage state estimation and LQR feedback control for voltage adjustment. The approach to obtain the optimal sensor selection sequence is studied. A sub-optimal sequence is obtained by applying the \emph{sliding window algorithm}.
Simulation results show that the proposed sensor selection strategy achieves a 40\% performance gain over the baseline algorithm of the round-robin sensor polling.

\end{abstract}

\section{Introduction and Motivation}
	  Ordinary electrical grids in many countries are undergoing a revolutionary change of evolving to smart grids, which are characterized by a two-way flow of electricity and information and will be capable of monitoring everything in the grid \cite{Smartgrid2008}. By bringing in a variety of Distributed Energy Resources (DERs), in particular renewable sources such as solar panels and wind turbines, smart grid addresses both globe warming and emergency resilience issues.

\begin{figure}[t]
      \begin{center}
	\includegraphics[scale=0.45]{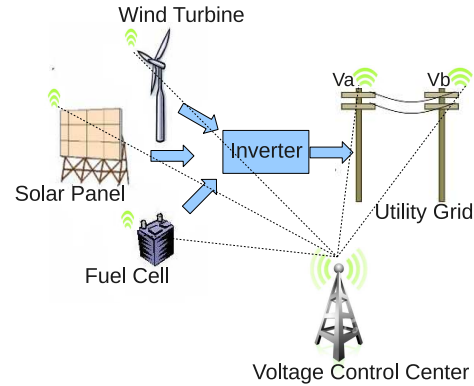}
      \end{center}
      \caption{\footnotesize{Distributed Energy Resources for Voltage Control}}
      \label{fig:Intro}
    \end{figure}

	In this paper we study {\em the wireless communication protocol design for regulating the voltages in smart grid which has a shared communication channel among control center, voltage sensors and DERs}. The feasibility of DER for regulating voltage has been well reported in the literature, such as \cite{Ko2007}. As shown in Fig. \ref{fig:Intro}, multiple sensors monitor the voltage states at $\{V_a, V_b,\dots\}$ and report the states to the voltage control center. Based on all received reports, the control center estimates the voltage states. If the estimated voltage state is deviated from a preset desired value, the control center coordinates all available DERs to regulate voltage. The arrival of new report from sensor triggers the control center to perform another round of voltage state estimation and regulating. The above iterative voltage regulating process continues until the the voltages are within a desired range.	

	The motivation of our work is the increasing availability of DERs, voltage sensors, and the overlaying communication network in power networks\cite{Vaccaro2010}. Although the voltage regulation for power system stability has been a critical problem under intensive study \cite{Jin2010,Kashem2005,Ko2007,Li2010}, the new properties and potential benefits brought by these new facilities still need more effort to uncover. Take the \emph{microgrid} as an example \cite{Hatz2007}, in which DERs have demonstrated their abilities of increasing power quality and reliability in practical systems; however, developing alternative control strategies using next-generation information and communication technology is still an open question. Existing solutions for the voltage control problem with DERs either focus on analyzing the power system model or mainly study control method design, e.g., the Model Predictive Control (MPC) \cite{Jin2010,Kashem2005} or PID controller \cite{Ko2007, Li2010}.

We will focus on {\em the MAC layer design of the wireless sensor networks for the voltage regulation}. When orthogonal communications are required, i.e., only one transmitter can access the channel at the same time and collisions incur packet loss\footnote{It is straightforward to extend to the case that multiple sensors can be scheduled simultaneously}, it is of key importance to study the problem of sensor selection. A simple solution is to use round-robin scheduling, i.e., the sensors take regular turns to report their measurements, regardless of the current voltage states. However, as will be seen later, a significant performance gain over the simple scheme will be achieved by manipulating the sensors in a system state aware manner. Particularly, we will model the power grid as a hybrid system \cite{Lunze2009}\cite{Savkin2002}, in which the power system is the continuous subsystem while the communication system is the discrete subsystem. The sensor selection will be considered as the switching of the system dynamics mode. Then, we apply a sliding window algorithm to optimize the sensor selection, or equivalently, the system mode selection. To our best knowledge, there have not been any studies applying the hybrid system theory to the communication protocol design in smart grid.

	The remainder of the paper is organized as follows. In Section \ref{sec:intro}, we introduce the modeling of power system with sensors reporting through a shared communication channel, as well as the formulation of the voltage control problem. In Section \ref{sec:algorithm}, we describe the voltage control procedures and the algorithm for optimizing the sensor selection in the MAC layer. An example application and its simulation results will be reported in Section \ref{sec:simulation}. The conclusions are given in Section \ref{sec:conclusion}.

\section{System Model and Problem Formulation}\label{sec:intro}
In this section, we first introduce the system model, including the power dynamics, system cost and communication system. Then, we formulate the problem as three sub-problems.

\subsection{Power System Dynamics}
We use the following differential-algebraic equation (DAE) to describe our target power system, which is given by
\begin{equation}
  \mathbf{\dot{x}}=f(\mathbf{x,u,w'}), \  \mathbf{w'}\thicksim \mathcal{N}(\mathbf{0}, Q'),
  \label{sysStat}
\end{equation}
where $\mathbf{x} \in \mathbb{R}^n$ is the system state representing the voltages; $\mathbf{u} \in \mathbb{R}^m$ is the system control action; $\mathbf{w'}$ is the system process noise which is assumed to be zero-mean, Gaussian and white with covariance matrix $Q'$. Since the voltage is usually required to stay within a narrow range centered at a desired value, we assume that the function $f$ can be well approximated by its linearization in the neighborhood of desired voltage values.

When voltage fluctuations due to either fault in the system or load change, DERs are able to provide compensation to regulate the voltages. Note that the action taken by a DER, say increasing its voltage, can affect all voltages in the system, more or less. Hence, a single DER as an individual actuator cannot reduce the voltage oscillation efficiently, as it does not have the global information on the voltage state. Therefore, to enable the DERs to collaboratively regulate the power system voltages, we must first obtain as much information as possible about the overall voltage state, and then assign the tasks of voltage adjustment to each DER accordingly.

There exists a group of sensors, ${S_1,\dots,S_i,\dots,S_N}$, monitoring voltage change in the power system. Each is able to obtain a partial observation of the system with its unique measuring function, which is given by
\begin{equation}
  \mathbf{y_{i}}=h_i(\mathbf{x,v'_i}), \ \mathbf{v'_i}\thicksim \mathcal{N}(\mathbf{0}, R'_i),\ i=1,\dots, N,
  \label{meas}
\end{equation}
where $\mathbf{y_{i}}$ denotes the measurement obtained by sensor $S_i$; $h_i$ is the measuring function associated with $S_i$; $\mathbf{v'_i}$ is the Gaussian measurement noise with zero mean and covariance matrix $R'_i$. We assume that $\mathbf{v'_i}$ is independent of the system process noise $\mathbf{w'}$.

With the controlled voltage staying close to preset desired value, we consider a discrete-time linearized model derived from aforementioned DAE. The time continuous functions $f$ and $h_i$ are locally linearized around desired voltage $\mathbf{x^*}$, which are given by
\begin{equation}
  \begin{split}
  \mathbf{x}_k=& f(\mathbf{x}_{k-1}^{*},\mathbf{u}_{k-1}^{*},\mathbf{0})+A(\mathbf{x}_{k-1}-\mathbf{x}_{k-1}^{*})\\
  		& +B(\mathbf{u}_{k-1}-\mathbf{u}_{k-1}^{*})+F\mathbf{w}'.
\end{split}
  \label{eq:sysLinear1}
\end{equation}
and
\begin{equation}
  \mathbf{y}_{ik}=h_i(\mathbf{x}_k^*,\mathbf{0})+H_{ik}(\mathbf{x}_k-\mathbf{x}_k^*)+G_i\mathbf{v}'_i,\ i=1, \dots, N,
  \label{eq:measLinear1}
\end{equation}
where $A, B, F, H_{ik} \ \mbox{and}\ G_i$ are matrices derived from the Jacobian matrices of $f$ and $h_i$; $\mathbf{x}_{k-1}^*=\mathbf{x}_k^*=\mathbf{x}^*$. Calculation of the Jacobian matrices and discrete-continuous model conversion are standard procedures \cite{Negenborn2007}\cite{Shieh1980}.
Since at the steady state, the voltages stay at desired values and the control action is not needed, we have $\mathbf{u}_{k-1}^{*}=\mathbf{0}\ \mbox{and}\ f(\mathbf{x}_{k-1}^{*}, \mathbf{u}_{k-1}^{*},\mathbf{0})=\mathbf{x}^{*}$, $h_i(\mathbf{x}_k^*,\mathbf{0})=\mathbf{y}_i^*$.
Substitute them into Eq. (\ref{eq:sysLinear1}) and Eq.(\ref{eq:measLinear1}) respectively, we have
\begin{equation}
  \mathbf{x}_k=\mathbf{x}^*+A(\mathbf{x}_{k-1}-\mathbf{x}^*)+B\mathbf{u}_{k-1}+F\mathbf{w}',
  \label{eq:sysLinear2}
\end{equation}
and
\begin{equation}
  \mathbf{y}_{ik}=\mathbf{y}_i^*+H_{ik}(\mathbf{x}_k-\mathbf{x}^*)+G_i\mathbf{v}'_i, \ i=1, \dots, N.
  \label{eq:measLinear2}
\end{equation}
Letting $\Delta \mathbf{x}_k=\mathbf{x}_k-\mathbf{x}^*$, $\Delta \mathbf{y}_{ik}=\mathbf{y}_{ik}-\mathbf{y}_i^*$, $\mathbf{w}=F\mathbf{w}'$ and $\mathbf{v}_i=G\mathbf{v}'_i$, we obtain the voltage deviation based system equation (\ref{eq:sysDelta}) and the measurement equation (\ref{eq:measDelta}), which are given by
\begin{equation}
  \Delta \mathbf{x}_k=A \Delta \mathbf{x}_{k-1}+ B\mathbf{u}_{k-1}+\mathbf{w}, \ \mathbf{w} \thicksim \mathcal{N}(\mathbf{0}, Q),
  \label{eq:sysDelta}
\end{equation}
\begin{equation}
  \Delta \mathbf{y}_{ik}=H_{ik} \Delta \mathbf{x}_k+ \mathbf{v}_i, \ \mathbf{v}_i \thicksim \mathcal{N}(\mathbf{0}, R_i), i=1, \dots, N,
  \label{eq:measDelta}
\end{equation}
where $Q=FQ'F^T$, $R_i=G_iR'_iG_i^T$. $Q$ represents the power system uncertainties which may be due to variations in the power system parameters, the effects of nonlinearities and the dynamics that have not been included in the power system model. $R_i$ reflects the uncertainties of sensor $i$'s measurement mainly because of noise.

\subsection{System Cost}
We define the time discretized cost function for the system as a quadratic function which penalizes the voltage deviation and minimizes control cost, which is given by
\begin{equation}
  J=\mathbb{E}(\sum_{k=1}^{k=K}(\Delta \mathbf{x}_k^TD\Delta \mathbf{x}_k+\mathbf{u}_k^TE\mathbf{u}_k)),
  \label{eq:cost}
\end{equation}
in which, $k=1\thicksim K$ is the entire voltage adjusting period. $D$ and $E$ are positive definite matrices whose weighting elements depend on power system's penalties for voltage deviations at different buses and different DERs' operating costs.

\subsection{Communication System}
We assume that the sensors can report their measurements to the control center equipped with a base station. The center can then compute the corresponding actions and send them to the DERs. Due to the expensive cost of wired communications, we assume that wireless communication technologies are employed. To avoid the possible collisions, the reports from the sensors are conveyed in a polling manner, i.e., the control center schedules the transmission of the sensors. For simplicity, we assume that only one sensor can be scheduled in a time slot and it is straightforward to extend to the case of multiple scheduled sensors. Moreover, we ignore the communication details like modulation and coding, as well as the transmission delay and packet drops, thus focusing on the sensor selection in the MAC layer.

\subsection{Problem Formulation}\label{subsec:subprob}
Our focus is to find an effective algorithm for selecting the voltage sensors. To that end, three subproblems have to be studied towards solving our problem of timely regulating voltage with minimum operating cost: \romannumeral 1) how to obtain the optimal system state estimation with partial observation from chosen sensors; \romannumeral 2) what control method should be applied based on the estimated system state; \romannumeral 3) which sensor to choose at each time slot and what is the selection criterion.

\section{Optimal Sensor Selection Sequence}\label{sec:algorithm}
In this section, we present our algorithm of sensor selection for the voltage control by employing the framework of hybrid dynamical systems \cite{Savkin2002}. We will first introduce the theory of hybrid dynamical systems. Then, we will explain the algorithm of sensor selection. 

\subsection{Hybrid Dynamical System}
Hybrid dynamical system (HDS) is a dynamical system which consists of both discrete and continuous dynamics. While continuous dynamics come from continuous subsystems of HDS, discrete dynamics are from the switching among these subsystems. Thus, the interaction between the discrete and continuous dynamics is the focus of HDS study.

One well known method of describing hybrid dynamical systems is using a set of ordinary differential equations with the following format:
\begin{equation}
  \dot{\mathbf{x}}(t)=f_i(\mathbf{x})
  \label{eq:hds}
\end{equation}
in which, $\mathbf{x}(t)\in \mathbb{R}^n$ is the system state; $i={1,2,\dots,N}$ is the switching system mode, and $f_1, f_2,\dots, f_N$ are continuous functions determined by the corresponding subsystems in the HDS.

Most dynamical systems around us are hybrid dynamical systems. Especially with advancement of modern digital technology, numerous systems have been equipped with computer based controller with digital-sampling blocks, which inevitably changes these systems into HDS. One example of such HDS is robotic system. It uses camera or other sensors to monitor surrounding environment, and chooses the optimal operating mode accordingly. Being a practical analysis model for a variety of modern systems, HDS has received intensive studies in the literature \cite{Labinaz1997,Schaft1998,Seatzu2006}.

\subsection{Sensor Selection Algorithm}
The power voltage control system under our study belongs to an important class of hybrid dynamical system called switching system, in which the continuous variables are the state variables of all continuous time subsystems and the discrete variables are the indices of subsystems. Specifically, in our power system, the continuous variables are the voltage states while the discrete variables are the indices of the chosen sensors.

We use feedback control to regulate the voltage. Since at any given time slot only one sensor can report, the power system is always under partial observations. To perform the feedback control, an estimation of overall voltage state has to be obtained first. The feedback control equation is given by
\begin{equation}
  \mathbf{u}_k=-L_k\times\hat{\mathbf{x}}_k=-L_k\times g(\mathbf{y}_{ik}),
  \label{eq:ctrlInput}
\end{equation}
where $\mathbf{u}_k$ is the control input; $L_k$ is the feedback control matrix; $\hat{\mathbf{x}}_k=g(\mathbf{y}_{ik})$ is the state estimation based on sensor $i$'s measurement $\mathbf{y}_{ik}$ and previously received measurements; $g(\cdot)$ is the estimation function. From (\ref{eq:sysDelta}), (\ref{eq:measDelta}) and (\ref{eq:ctrlInput}), we have
\begin{equation}
  \Delta \mathbf{x}_{k+1}=A\times \Delta \mathbf{x}_k- B\times L_k\times g(H_{ik}\Delta \mathbf{x}_k+\mathbf{v}_i)+\mathbf{w}.
  \label{eq:feedback}
\end{equation}
With Eq. (\ref{eq:feedback}), we revisit the three subproblems (section \ref{subsec:subprob}) to be solved for achieving our ultimate goal of sensor selection in voltage control: \romannumeral 1) function $g(\cdot)$ gives system state estimation; here we use Kalman filter; \romannumeral 2) feedback matrix $L_k$ represents control method for which we adopt Linear Quadratic Regulator (LQR); \romannumeral 3) $H_{ik}$ indicates the choice among different sensors. The following three subsections address these three individual problems.

\subsubsection{Kalman Filter for State Estimation}
The Kalman filter is a set of mathematical equations that provide an efficient recursive computational means to estimate the state of a process by minimizing the mean square error \cite{Welch1995}. The state estimation process has two main interactive procedures: process update and measurement update whose mathematical expressions for our specific voltage control problem are Eq. (\ref{eq:processUpdate}) and Eq.(\ref{eq:measUpdate}) respectively, namely
\begin{equation}
  \Delta \hat{\mathbf{x}}_k^-= A \Delta \hat{\mathbf{x}}_{k-1}+B\mathbf{u}_{k-1},
  \label{eq:processUpdate}
\end{equation}
and
\begin{equation}
  \Delta \hat{\mathbf{x}}_k=\Delta \hat{\mathbf{x}}_k^-+K_k(\mathbf{y}_{ik}-H_{ik}\Delta \hat{\mathbf{x}}_k^-),
  \label{eq:measUpdate}
\end{equation}
in which $\Delta \hat{\mathbf{x}}_k^-$ is the preliminary voltage deviation estimation based on the system state dynamics in (\ref{eq:sysDelta}) with control input applied; $\Delta \hat{\mathbf{x}}_k$ is the refined voltage deviation estimation after incorporating the correction provided by current measurement $\mathbf{y}_{ik}$; $K_k$ is the Kalman gain matrix which can be calculated beforehand according to Eq. (\ref{eq:kalmanGain}), namely
\begin{equation}
  K_k=P_k^-H_{ik}^T(H_{ik}P_k^-H_{ik}^T+R_i)^{-1},
  \label{eq:kalmanGain}
\end{equation}
where $P_k^-=AP_{k-1}A^T+Q$ is the predicted estimation covariance which is iteratively updated by $P_k=(I-K_kH_{ik})P_k^-$.

\subsubsection{LQR for Feedback Control}
With the latest state estimation available from Kalman filtering, we use Linear Quadratic Regulator (LQR) \cite{Sontag1998} to control the deviated voltage to the desired value. Being an effective control method in solving problem with linear system model and quadratic cost function, LQR is a good fit for voltage control. In fact, LQR, together with the Kalman filter, forms a Linear Quadratic Gaussian (LQG) problem. By LQG separation principle \cite{Zhang2005},
we are able to decouple the voltage state estimation from LQR control and calculate feedback matrix $L_k$ in advance by Eq. (\ref{eq:feedbackMatrix}), which avoids posing a substantial computation burden on voltage control center.
\begin{equation}
  L_k=(E+B^TM_kB)^{-1}B^TM_kA,
  \label{eq:feedbackMatrix}
\end{equation}
in which $A$ and $B$ are system matrices in Eq. (\ref{eq:sysDelta}); $E$ is the control input cost matrix in the cost function (\ref{eq:cost}); $M_k$ is found iteratively backwards in time by using the following equation:
\begin{equation}
  M_{k-1}=D+A^T(M_k-M_kB(E+B^TM_kB)^{-1}B^TM_k)A,
  \label{eq:lqrUpdate}
\end{equation}
with initial condition $M_K=D$, and $D$ is the voltage deviation cost matrix in the cost function (\ref{eq:cost}).

\subsubsection{Sensor Selection}
Now we face the key challenge of sensor selection. We denote the sensor querying sequence by $I=\{i_1,\dots,i_k,\dots,i_K\}$ for $k=1\thicksim K$, and $i_k \in \{1,\dots,N\}$. Since the measurement of the current selected sensor, together with all previous sensor reports, determines the voltage control input which in turn determines the voltage states, the system cost function (\ref{eq:cost}) becomes a function of $I$. Hence, our goal is to minimize the overall cost by finding an optimal sensor querying sequence $I$, i.e.,
\begin{equation}
  \mathop{min}_I\{J(I)=\mathbb{E}(\sum_{k=1}^{k=K}(\Delta \mathbf{x}_k^TD\Delta \mathbf{x}_k+\mathbf{u}_k^TE\mathbf{u}_k))\}.
  \label{eq:costFunOfI}
\end{equation}

According to the separation principle of LQG problem, its optimal control is totally based on the accurate state estimation. Therefore the optimal sensor querying sequence is the one that can achieve the minimum voltage deviation estimation error. The estimation error covariance is given by
\begin{equation}
  P_k=\mathbb{E}[(\Delta \mathbf{x}_k-\Delta \hat{\mathbf{x}}_k)(\Delta \mathbf{x}_k-\Delta \hat{\mathbf{x}}_k)^T].
  \label{eq:estError}
\end{equation}

From $k=1$ to $k=K$, our goal consequently becomes finding the optimal (or near optimal) sensor querying sequence $I$ which minimizes overall estimation error, which can be written as
\begin{equation}
  \mathop{min}_I\{\sum_{k=1}^{k=K}trace(P_k)\}.
  \label{eq:trace}
\end{equation}

By employing Eq. (\ref{eq:kalmanGain}) and the iterative updating process for Kalman gain $K_k$, the estimation error covariance evolves as follows:
\begin{eqnarray}
  P_k&=&[I-P_k^-H_{ik}^T(H_{ik}P_k^-H_{ik}^T+R_i)^{-1}H_{ik}]P_k^-,\\
  P_k^-&=&AP_{k-1}A^T+Q.
  \label{eq:evolve}
\end{eqnarray}
The initial value $P_0$ can be an approximate one which reflects estimation accuracy of given $\hat{\mathbf{x}}_0$.

Starting from the selection of sensor at $k=1$ until the voltage is adjusted to the desired value at $k=K$, we have $N$ choices in each step. Thus, we can grow a tree structure for all possible sensor querying sequences. To find the optimal sequence, one straightforward but inefficient method is the \emph{brute force strategy} which traverses all sequences and selects the one with the minimum estimation error as required by (\ref{eq:trace}). While it guarantees to find the optimal sequence, the \emph{brute force strategy} suffers the exponential increase of computational cost.

We seek a trade-off between the sub-optimality sensor sequence and reasonable computation effort by adopting the \emph{sliding window algorithm} \cite{Chung2004}. Given a window size $d$(steps), the algorithm proceeds as follows:
\begin{enumerate}
  \item \emph{Initialization}: start from root node with time $k=1$.
  \item \emph{Traversal}:
    \begin{enumerate}
      \item Traverse all the possible paths in the tree for the next $d$ levels from the present node;
      \item Identify the optimal sensor sequence within the $d$-window;
      \item Put the first sensor of the optimal sequence into the output sensor sequence.
    \end{enumerate}
  \item \emph{Sliding the window}:
    \begin{enumerate}
      \item If $k=K$ then quit, otherwise go to the next step;
      \item Use the sensor which has just been selected as the new root;
      \item Update time $k=k+1$;
      \item Repeat the traversal step.
    \end{enumerate}
\end{enumerate}
In the algorithm, the window size $d$ is an adjustable parameter determining the trade-off between the sequence optimality and computational cost (or the speed of decision making). Larger window size $d$ results in a better sensor sequence but more computational intensity, and vice versa. As pointed out in \cite{Chung2004}, when we slide the window, the first $d-1$ steps' error covariances in the new window have already been calculated in the previous window and are available for immediate use. This merit of the algorithm considerably reduces computational demand.

\section{Example Application and Simulation Results}\label{sec:simulation}
\begin{figure}[tb]
  \begin{center}
    \includegraphics[scale=0.4]{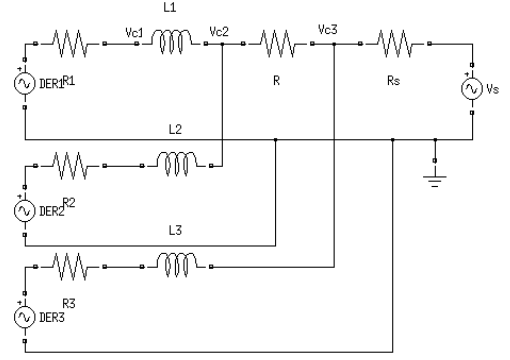}
  \end{center}
  \caption{\footnotesize{Example Power System Model}}
  \label{fig:model}
\end{figure}
In this section, we use an example application of power voltage control (Fig. \ref{fig:model}) to demonstrate the effectiveness of our proposed sensor selection strategy. In this example application, three voltage controlling/regulating DERs and three voltage monitoring sensors are installed in the power system. The system matrices in the power system equation (\ref{eq:sysDelta}) are give by
\begin{eqnarray}
A=\left(
\begin{array}{ccc}
1.03 & 0 & 0\\
0 & 1.02& 0\\
0 & 0& 1.05\\
\end{array}
\right),
\end{eqnarray}
and
\begin{eqnarray}
B=\left(
\begin{array}{ccc}
0.6 & 0.1&0.2\\
0.1 & 0.7&0.15\\
0.2&0.15&0.8\\
\end{array}
\right).
\end{eqnarray}
Elements in $A$ being larger than 1 means that, without timely curbing voltage deviated from desired value, the state in the system will keep deteriorating. $B$ shows that action of any single DER affects the state of the entire power system and DERs' control capabilities are coupled with one another, though each DER has its own primary control area.
The covariance matrices of the system process noise $w$ and the sensor measurement noise $v$ are given by
\begin{eqnarray}
Q=\left(
\begin{array}{ccc}
0.05 & 0 & 0\\
0 & 0.02& 0\\
0 & 0& 0.01\\
\end{array}
\right),
\end{eqnarray}
and
\begin{eqnarray}
R=\left(
\begin{array}{ccc}
0.1 & 0 & 0\\
0 & 0.2& 0\\
0 & 0& 2\\
\end{array}
\right).
\end{eqnarray}

The noise power at sensor 3 is set to be much larger than those at the other two sensors, because we want to show that our sensor selection strategy is able to compensate the inferior condition by optimally allocating the shared communication channel. We also give voltage deviation penalty matrix $D$ and control input (DER operation) cost matrix $E$ in (\ref{eq:cost}), which are given by
\begin{equation}
D=\left(
\begin{array}{ccc}
1 & 0 & 0\\
0 & 2& 0\\
0 & 0& 3\\
\end{array}
\right),
\mbox{and} \
E=\left(
\begin{array}{ccc}
5 & 0&0\\
0 & 5&0\\
0 & 0&5\\
\end{array}
\right).
\end{equation}

We set the initial voltage deviation as $\Delta \mathbf{x}_0=[30,10,20]^T$, and use two different methods to perform the sensor selection: one is using our proposed sensor selection strategy which returns the sensor querying sequence below; the other is to use the \emph{round-robin} polling \{2 3 1 2 3 1 \dots\}, and this method is used as our baseline. The sliding window size $d$ is set as $d=5$.

\emph{Sensor querying sequence:} \{1 2 3 3 1 3 2 3 1 3 1 3 2 1 3 3 1 2 3 1 3 1 2 3 1 3 1 3 2 1 3 3 1 2 3 1 3 1 2 3\}
\begin{table}[b]
  \centering
  \begin{tabular}{|c|c|c|c|}
    \hline
    Sensor&$S_1$&$S_2$&$S_3$\\
    \hline
    Allocated Slots&14&8&18\\
    \hline
    Percentage & 35\%&20\%&45\%\\
    \hline
  \end{tabular}
  \caption[belowskip=-10pt]{\footnotesize{Communication Channel Allocation}}
  \label{tab:alloc}
\end{table}

Table \ref{tab:alloc} gives the communication channel allocation statistics for all three sensors. Sensor 3 receives the highest utilization percentage of the channel, i.e., 45\% of channel accesses, while sensor 1 and sensor 2 receive 35\% and 20\% respectively. According to the noise covariance matrices $Q$ and $R$, sensor 3 suffers the highest level of measurement noise which is overwhelming compared with the other two's; thus it is granted the highest utilization percentage. Sensor 1 has a lower measurement noise but higher process noise than sensor 2; consequently, sensor 1's combined noise effect gives it larger channel utilization percentage (35\%) than sensor 2 receives (20\%).

\begin{figure}[tb]
  \begin{center}
    \includegraphics[scale=0.5]{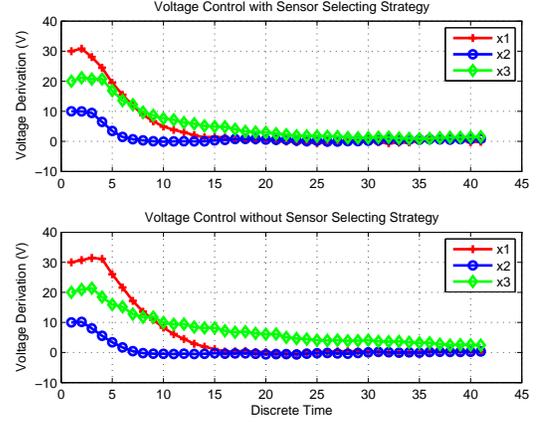}
  \end{center}
  \caption{\footnotesize{Voltage State Evolution. Controlled by sensor selection strategy, deviation of voltage is eliminated by the time $k=30$; Without the strategy, deviation still exists after $k=40$.}}
  \label{fig:volState}
\end{figure}

Fig. \ref{fig:volState} depicts the voltage state of the system during the control process using both methods. The lower figure uses the round-robin polling, and the upper figure shows results using our proposed strategy. Both methods successfully pull the deviated voltage back to the desired value (deviation becomes zero), while our method forces the voltage to converge faster, in particular for voltage 3 (the green line).

\begin{figure}[tb]
  \begin{center}
    \includegraphics[scale=0.5]{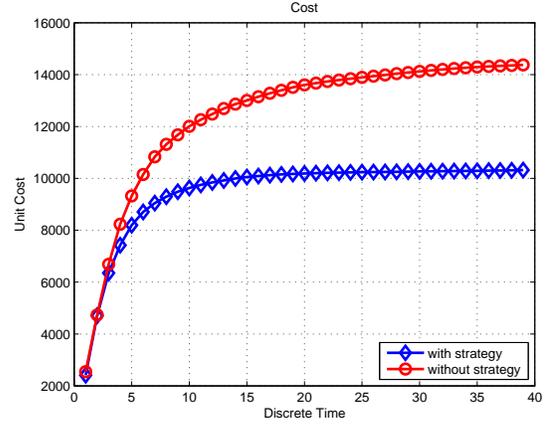}
  \end{center}
  \caption{\footnotesize{Cost Comparison. The method with sensor selection strategy reduces cost by approximately 40\% compared with the method using \emph{round-robin} sensor polling}}
  \label{fig:cost}
\end{figure}

Approaching the desired voltage $\mathbf{x}^*$ faster results in less time staying deviated from $\mathbf{x}^*$ and thus reduces the cost. Fig. \ref{fig:cost} shows the costs for both voltage control methods. Our proposed strategy outperforms the baseline of round-robin algorithm by reducing the cost by approximately 40\%.

\begin{figure}[tb]
  \begin{center}
    \includegraphics[scale=0.5]{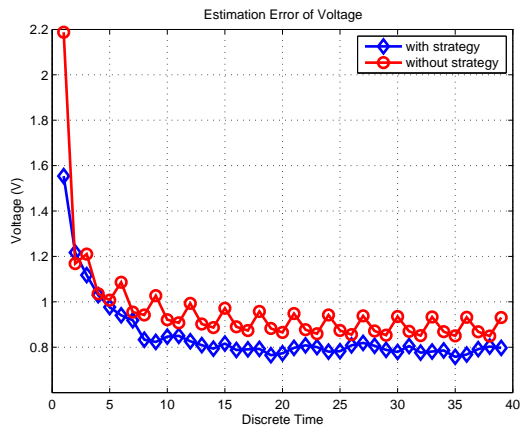}
  \end{center}
  \caption{\footnotesize{State Estimation Error.}}
  \label{fig:EstError}
\end{figure}

\begin{figure}[tb]
  \begin{center}
    \includegraphics[scale=0.60]{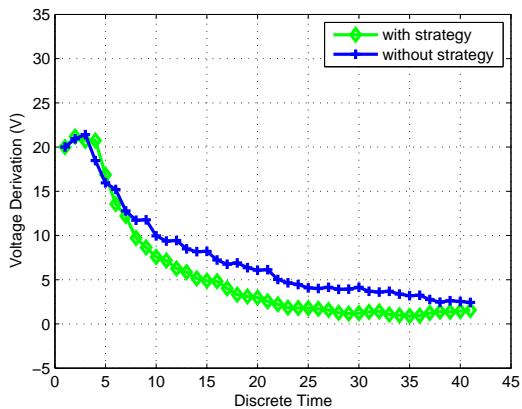}
  \end{center}
  \caption{\footnotesize{Voltage State Transition. \emph{round-robin} sensor polling results in moderate state fluctuation (the upper curve); voltage state controlled by sensor selection strategy has smoother transition (the lower curve).}}
  \label{fig:conv}
\end{figure}

One of the key reasons that our sensor selection method is able to beat the round-robin method is that our method achieves smaller voltage state estimation error, as demonstrated in Fig. \ref{fig:EstError}. The more accurate state estimation of our proposed algorithm helps the control center to timely use DERs to adjust voltage states and reduce the state fluctuation. In Fig. \ref{fig:conv}, the voltage state controlled by the round-robin method, namely the upper curve, has moderate fluctuation from time $k=5$ to $k=25$; while the voltage state controlled by our sensor selection strategy, the lower curve, shows smooth transition. Furthermore, the voltage state with a smoother transition like the one controlled by our sensor selection strategy is much more desired.

\section{Conclusion and Future Work}\label{sec:conclusion}
In this paper, we have treated the power system with sensors and shared communication channel as a hybrid dynamical system, which switches its mode by selecting different sensors. The approach to obtain the optimal sensor querying sequence has been analyzed by minimizing overall system cost. Both LQR control and Kalman filter have been applied for the control. A sub-optimal but computational efficient sliding window algorithm has been applied and has been demonstrated to achieve a 40\% performance gain compared with the simple round-robin sensor polling baseline. 

We did not consider more details of the communication system, e.g., the delay and packet drop rate. In the future, we will study how these factors affect the system performance of voltage control. Moreover, we assume a base station is available for the information processing and communication scheduling. In practical system like microgrid, it may be more desirable to eliminate the necessity of such a base station due to its expensive cost. Hence, we will study the sensor scheduling in a complete decentralized communication infrastructure in smart grid.


\begin{thebibliography}{18}
  \bibitem{Vaccaro2010}
    A. Vaccaro, D. Villacci, M. Osborne, J. Fitch, D. Cai; V. Terzija, ``The role of cooperative sensor networks in wide area power systems communication,'' \emph{Developments in Power System Protection (DPSP 2010)}, pp.1-5, Mar., 2010.
    
  \bibitem{Smartgrid2008}
   Department of Energy's Office of Electricity Delivery and Energy Reliability, ``The smart grid: an introduction,'' 2008.
  
\bibitem{Hatz2007}
	N. Hatziargyriou, H. Asano, R. Iravani and C. Marnay, ``Microgrids: an overview of ongoing research, development, and demonstration projects,''  \emph{IEEE Power Energy Mag.},  vol. 5,  pp. 78, 2007.
      \bibitem{Jin2010}
	L. Jin, R. Kumar and N. Elia, ``Model predictive control-based real-time power system protection schemes,'' \emph{IEEE Trans. on Power Systems}, vol. 25, no. 2, May, 2010.
      \bibitem{Kashem2005}
	M.A. Kashem and G. Ledwich, ``Multiple distributed generators for distributed feeder voltage support,'' \emph{IEEE Trans. on Energy Conversion}, vol. 20, no. 3, Sep., 2005.
      \bibitem{Ko2007}
	H. Ko, G. Yoon and W. Hong, ``Active use of DFIG-based variable-speed wind-turbine for voltage regulation at a remote location,'' \emph{IEEE Trans. on Power Systems}, vol. 22, no. 4, pp. 1916-1925, Nov., 2007.
      \bibitem{Li2010}
	H. Li, F. Li, Y. Xu, D.T. Rizy and J.D. Kueck, ``Adaptive voltage control with distributed energy resources: Algorithm, theoretical analysis, simulation, and field test verification,'' \emph{IEEE Trans. on Power Systems}, vol. 25, no. 3, Aug., 2010.
      \bibitem{Lunze2009}
	J. Lunze and F. L. Lagarrigue, edit, \emph{Handbook of Hybrid Systems Control: Theory, Tools, Applications}, Cambridge University Press, 2009.
  \bibitem{Negenborn2007}
    R.R. Negenborn, A.G. Beccuti, T. Demiray, S. Leirens, G. Damm, B. De Schutter and M. Morari, ``Supervisory hybrid model predictive control for voltage stability of power networks,'' \emph{Proceedings of the 2007 American Control Conference}, New York, New York, pp. 5444-5449, Jul. 2007.
  \bibitem{Shieh1980}
    L.S. Shieh, H. Wang, R.E. Yates, ``Discrete-continuous model conversion,'' \emph{Applied Mathematical Modelling}, Vol.4, Issue 6, pp.449-455, Dec. 1980.
  \bibitem{Savkin2002}
    A.V. Savkin and R.J. Evans, {\em Hybrid Dynamical Systems: Controller and Sensor Switching Problems,} Boston, MA: Birkh$\ddot{a}$user, 2002.
  \bibitem{Zhang2005}
    L. Zhang, D. Hristu-Varsakelis, ``LQG control under limited communication,'' \emph{Decision and Control, 2005 and 2005 European Control Conference. CDC-ECC '05, 44th IEEE Conference on}, pp. 185- 190, Dec. 2005.
  \bibitem{Welch1995}
    G. Welch and G. Bishop, ``An introduction to the Kalman filter,'' \emph{Technical Report TR 95-041}, University of North Carolina at Chapel Hill, 1995.
  \bibitem{Sontag1998}
    E. Sontag, \emph{Mathematical Control Theory: Deterministic Finite Dimensional Systems}, second edition, Springer, 1998.
  \bibitem{Labinaz1997}
    G. Labinaz and M. M. Bayoumi and K. Rudie, ``A survey of modeling and control of hybrid systems,'' \emph{Annual Reviews of Control}, vol. 21, pp. 79--92, 1997.
  \bibitem{Schaft1998}
    A. van der Schaft and H. Schumacher. ``Complementarity modeling of hybrid systems,'' \emph{IEEE Transactions on Automatic Control,} 43(4):483-490, 1998.
  \bibitem{Seatzu2006}
    C. Seatzu, D. Corona, A. Gina and A. Bempord, ``Optimal control of continuous-time switched affine systems,'' \emph{IEEE Transactions On Automatic Control}, vol. 51, no.5, May, 2006.
  \bibitem{Chung2004}
    T. H. Chung, V. Gupta, B. Hassibi, J. Burdick and R. M. Murray, ``Scheduling for distributed sensor networks with single sensor measurement per time step,'' \emph{Proceedings of the 2004 IEEE International Conference on Robotics and Automation}, New Orleans, LA, Apr., 2004.

\end{thebibliography}
\end{document}